\documentclass[prd,nofootinbib,
amsmath,amssymb,aps,
reprint,
floatfix
]{revtex4-2}

\usepackage[colorlinks=true
,urlcolor=blue
,anchorcolor=blue
,citecolor=blue
,filecolor=blue
,linkcolor=blue
,menucolor=blue
,linktocpage=true
,pdfproducer=medialab
,pdfa=true
]{hyperref}

\usepackage{hyperref}
\usepackage[T1]{fontenc}
\usepackage{graphicx}
\usepackage{comment}
\usepackage{bm}
\usepackage{braket}
\usepackage{cancel}
\usepackage{lmodern}
\usepackage{ulem}
\usepackage{amsmath}

\newcommand\apjl{ApJ}                
\newcommand\apjs{ApJS}               
\newcommand\jcap{J.~Cosmology Astropart. Phys.} 
\newcommand\mnras{MNRAS}             
\newcommand\physrep{Phys.~Rep.}      

\newcommand{\eq}[1]{\begin{equation}\begin{split} #1 \end{split}\end{equation}}

\newcommand{\lr}[1]{\left( #1 \right)}

\newcommand{\m}{\mathrm}

\begin{document}

\title{Cosmological contribution from population III stars in ultracompact minihalos}
\author{Katsuya T. Abe}
\email{abe.katsuya.f3@s.mail.nagoya-u.ac.jp}
\affiliation{Division of Particle and Astrophysical Science,
Graduate School of Science, Nagoya University,
Chikusa, Nagoya 464-8602, Japan}

\begin{abstract}
In this work, we investigate the effect of Population~III~(Pop.~III) stars in ultracompact minihalos~(UCMHs) on the cosmic ionization history using the Planck observation data.
Although high-redshift astrophysics is not understood yet, UCMHs could host the Pop.~III stars like the halos formed in the standard structure formation scenario. Such Pop.~III stars would emit ionizing photons during their main sequence and facilitate cosmic reionization in high redshifts. To study their effects on the global ionization, we model the cosmic reionization evolution based on the ``tanh"-type reionization model which is expressed by $z_{\m{reio}}$ with additional two parameters characterizing the initial mass of UCMHs and the number density of UCMHs.
We implement the Monte Carlo Markov Chain analysis with the latest Planck observation data for our reionization model. As the result, we found that if the UCMH initial mass is larger than $10^{8.4}\m{M}_{\odot}$, the number density of UCMHs is strictly limited. 
Then we obtained the constraint on the amplitude of the primordial power spectrum through the constraint on the UCMH number density like $\mathcal{A}_{\zeta}\lesssim 10^{-8}$ in the scales, $k\lesssim 50\m{Mpc}^{-1}$, when we assume that the standard ``tanh"-type reionization occurs by $z=3$, so that we set $z_{\m{reio}}>3$.
\end{abstract}

\maketitle

\section{Introduction}
Ultracompact minihalos~(UCMHs) are gravitational objects formed by denser regions of matter induced by the excess power of primordial scalar perturbations on small scales~\cite{2009ApJ...707..979R}.
Although there is no smoking gun observed events to detect the existence of UCMHs, they can be a strong cosmological probe of primordial scalar perturbation, especially in smaller scales, $k\gtrsim 1\m{Mpc}^{-1}$.
Since the dark matter density fluctuations can grow after entering the horizon even in the radiation-dominated epoch,
it would be possible to form minihalos in the early Universe like $z \sim 1000$ as long as the initial density perturbation is large enough at the horizon entry. That is the basic UCMH formation process.
Reference~\cite{2009ApJ...707..979R} theoretically suggests that UCMHs have a more compact profile with a larger central density than typical dark matter halos called NFW profile~\cite{10.1093/mnras/275.3.720} through the radial infalling in high redshift.
After that, Refs.~\cite{2018PhRvD..97d1303D,Delos+:2018a} performed the cosmological simulation of the UCMH formation for the spike-type power spectrum on small scales. 
They showed that the excess power of small-scale primordial scalar perturbation actually leads to the early structure formation, and formed UCMHs have the Moore-type matter density profile, $\rho \propto r^{-3/2}$ at the inner cusp region~\cite{1999MNRAS.310.1147M} which is steeper than the NFW profile.

If the dark matter is particle-type, especially the weakly-interacting massive particles~(WIMPs)~\citep{1985NuPhB.253..375S,1996PhR...267..195J,1999dmap.conf..592K},
UCMHs would emit energetic emissions through the WIMP annihilation enhanced by their dense profile.
So far, Refs.~\cite{2010PhRvD..82h3527J, 2009PhRvL.103u1301S, 2012PhRvD..85l5027B,
2016MNRAS.456.1394C,2018PhRvD..97b3539N,2018PhRvD..97d1303D,Delos+:2018a} have investigated the gamma-ray emission signal and provided the constraint on the UCMH abundance. Then they also provided the constraint on the small-scale primordial scalar perturbation, $\mathcal{A}_{\zeta} < 10^{-7}$ for $10~{\rm Mpc}^{-1} <k < 10^8~{\rm Mpc}^{-1}$, through the nondetection of such energetic signals in gamma-ray observation like Fermi-LAT~\cite{Atwood_2009}.
In addition, UCMHs have been investigated through their gravitational lensing effect~\cite{2012PhRvD..86d3519L,2016MNRAS.456.1394C} and their contribution to the cosmic reionization~\cite{2011MNRAS.418.1850Z,2011JCAP...12..020Y,2016EPJP..131..432Y,2017JCAP...05..048C}.

The author focuses on the baryon gas within UCMHs. 
When the mass of UCMHs is larger than the Jeans mass, UCMHs would host the baryon gas.
Then through investigations in the cosmological signals from baryon gas within UCMHs, we can put constraints on the UCMH abundance and the small-scale primordial perturbation without the assumption of the dark matter nature.
In our previous work~\cite{2020MNRAS.494.4334F}, we investigated the 21cm line-emission anisotropy induced by UCMHs and provided the constraint on the primordial scalar perturbation on small scales, ${\mathcal A}_{\zeta} \lesssim 10^{-6}$ on $100~{\rm
Mpc}^{-1}\lesssim k \lesssim 1000~{\rm Mpc}^{-1}$.
Additionally in Ref.~\cite{2022PhRvD.105f3531A}, we have studied the free-free emission from UCMHs, and using the Planck free-free emission measurement, we have provided the constraint on the primordial scalar perturbation, $\mathcal{A}_\zeta \lesssim 10^{-7}$
for $1~\mathrm{Mpc}^{-1} \lesssim k \lesssim 100~\mathrm{Mpc}^{-1}$.

This paper shows another observable which has the potential to be much more constraining than the previous works.
We focus on astrophysical effects that occurred in UCMHs.
Since UCMHs could be formed in the early Universe and have a dense matter profile like the Moore profile, they might lead to the formation of stars at much higher redshifts than expected in standard cosmology. 
In that case, zero metallicity stars often called Population~III~(Pop.~III) stars would be formed from the primordial baryon gas mostly composed of hydrogen within UCMHs. 
According to the theoretical and numerical studies~\cite{2006ApJ...639..621A,2007ApJ...665...85J}, Pop.~III stars are considered to emit ionizing photons well.
Then the Pop.~III stars hosted by UCMHs would proceed with the ionization globally in high redshifts and modify the standard cosmic reionization history.
In Ref.~\cite{2020PhRvD.101l3518I}, they have studied similar astrophysical effects in axion clusters formed by isocurvature fluctuations. Although we are interested in the effects that occurred in UCMHs formed by the spike-type power spectrum rather than isocurvature-type, we refer to this work in our calculation. 


In the standard cosmology, after the cosmic recombination epoch at $z\sim 1000$, the kinematic decoupling of the baryons from the radiation occurs~\cite{1968ApJ...153....1P,1968ZhETF..55..278Z}, and the Universe enters the next epoch called the \textit{dark age} where the global ionization fraction is very low, $x_{\m{e}}\sim 10^{-4}$. This epoch will reach the end through the formations of stars, galaxies, etc, and their emission of ionizing photons; this is the standard cosmic dawn and reionization scenario.
We believe that the cosmic reionization has been mainly proceeded by Population~II (Pop.~II) stars and first galaxies~\cite{2022MNRAS.511.3657M} and almost completed by $z\sim 6$ as the observations of Lyman-$\alpha$~(Ly$\alpha$) absorption lines imprinted on quasar spectra indicate~\cite{2006AJ....132..117F}.
The luminosity of Ly$\alpha$ is quite sensitive to the amount of neutral hydrogen.
Therefore, observation of the redshift evolution of the Ly$\alpha$ luminosity can also provide significant information for the evolution of neutral hydrogen fractions during the epoch of reionization.
From the current observation of Ly$\alpha$ emitters, it is suggested that the number density of Ly$\alpha$ emitters decreases as the redshift increases, and the neutral hydrogen fraction increases from $z\sim 6-7$~\cite{Hu_2010,Ota_2010,2010ApJ...723..869O,2011ApJ...734..119K,2014ApJ...797...16K,2016MNRAS.463.1678S}.

One of the observables for the reionization history is CMB anisotropies. 
Especially the CMB E-mode polarization anisotropy is sensitive to the high redshift reionization expected by the UCMH Pop.~III stars model through the Compton scattering.
In this work, we investigate this model characterized by the mass and the number density of UCMHs using Markov Chain Monte Carlo~(MCMC) methods with the Planck 2018 observation likelihoods~\cite{Planck2018_cospara}.

This paper is organized as follows. 
In Sec.~II, we describe the properties of UCMHs formed by the spike-type curvature power spectrum.
In Sec.~III, we introduce the effect of Pop.~III stars formed in UCMHs on the global ionization fraction.
After that in Sec.~IV, we explain our cosmic reionization model considered here and show the relation between the reionization models and the anisotropies of the CMB temperature and E-mode polarization.
In Sec.~V, we explain the MCMC methods used in this work and show the results.
We also discuss the constraint on the amplitude of the primordial power spectrum suggested by the MCMC results.
Finally, we summarize in Sec.~VI.
Throughout our paper, we take the flat $\Lambda$CDM model with the Planck best-fit parameters obtained from TT, TE, EE, and low-$\ell$ + lensing observation data~\cite{Planck2018_cospara}.

\section{The properties of UCMHs}
The larger amplitude of the small-scale matter density fluctuation would provide the formation of the ultracompact minihalos~(UCMHs).
In this work, we assume the existence of the spike-type power spectrum on a specific small-scale $k_{\m{s}}$
in addition to the almost scale-invariant spectrum with amplitude $\mathcal{A}_{\zeta}^{\m{CMB}}\sim 2\times 10^{-9}$ measured by the Planck CMB observation.
Here, for simplicity, we adopt the Dirac delta function to represent the additional spike-type power spectrum as,
\eq{\label{eq: add_spike_pk}
\mathcal{P}^{\m{add}}_{\zeta}(k)=\mathcal{A}_{\zeta}^{\m{add}} k_{\rm s} \delta(k-k_{\rm s}),
}
where $\mathcal{A}_{\zeta}^{\m{add}}$ is the amplitude of the additional power spectrum, and $k_{\rm s}$ is the wave number corresponding to the spike center.
We begin with a brief summary of the features of UCMHs with the spike-type power spectrum~(see Refs.~\cite{2018PhRvD..97d1303D,Delos+:2018a} for details.)

In the spike-type power spectrum case, the initial mass of UCMHs $M_{\m{i}}$ are related to the spike-wave number $k_{\m{s}}$ like
\eq{\label{eq: ini_mass_ucmh}
M_{\m{i}} \sim 4 \times 10^{4}~{\rm M_{\odot}} \times \left(\frac{k_{\rm s}}{10^{3} ~ {\rm Mpc}^{-1}} \right)^{-3}.
}
The mass increase after their formation at redshift $z_{\m{f}}$ through late-time accretions from their outer region, e.g. intergalactic medium~(IGM), as~\cite{Delos+:2018a,2020MNRAS.494.4334F}
\eq{\label{eq: mass_evo_ucmh}
M(z,z_{\m{f}})  = M_{\m{i}}\lr{1+ \ln  \left( \frac{1+z_{\rm f}}{1+z} \right)}.
}
We assume that this accretion does not halt in our calculated duration, although Ref.~\cite{Delos+:2018a} suggests that this logarithmic mass increment does not necessarily continue to later times.
We also comment that in the radial infall theory for the mass accretion, the mass of UCMHs would grow like $M\propto a$~\cite{1985ApJS...58...39B} as several previous works (e.g. Ref.~\cite{2012PhRvD..85l5027B}) assumed. 
However, since this theory assumes that the existence of an overdense region in an unperturbed background region, which is unlike in the real Universe, we deal with the mass increase like Eq.~\eqref{eq: mass_evo_ucmh}.

The number density of UCMHs can be evaluated by employing the peak theory which is Bardeen, Bond, Kaiser, and Szalay (BBKS)~\cite{1986ApJ...304...15B} formulated in the spike-type power spectrum case.
This is because UCMHs form at the peak locations of the density fluctuations following the peak-type power spectrum of Eq.~\eqref{eq: add_spike_pk}.
Following the BBKS, the UCMH number density can be obtained by
\eq{\label{eq: num_dens_ucmh}
n(M_{\m{i}},z)=\frac{k_{s}^{3}}{(2 \pi)^{2} 3^{3 / 2}} \int_{\delta_{c}/S_{\m{mat,0}}^{1/2}(M_{\m{i}})D(z)}^{\infty} e^{-\nu^{2} / 2} f(\nu) \mathrm{d} \nu
}
where $\delta_{\rm c}=1.686$ is the linear density threshold for collapse, $S_{\m{mat,0}}$ is the present mass variance of the matter density fluctuation, $D(z)$ is the growth rate of matter density fluctuation which is $D(z)=(1+z)^{-1}$ during matter dominated epoch, and $f(\nu)$ is the function provided by equation~(A15) in the BBKS. It is noted that this expression of the number density in Eq.~\eqref{eq: num_dens_ucmh} does not include any merger effect about UCMHs, and we neglect them in this work.

Let us see the relation of the additional spike-type power spectrum in Eq.~\eqref{eq: add_spike_pk} and the present mass variance $S_{\m{mat,0}}$.
Basically, $S_{\m{mat,0}}$ is calculated from the power spectrum of the primordial curvature perturbations $\mathcal{P}_{\zeta}(k)$ by
\eq{
S&_{\m{mat,0}}(M) \\
&\ ~~~= \int d\log k ~\cfrac{4}{25}~\cfrac{k^4}{\Omega_\mathrm{m,0}^2 H_0^4} ~\mathcal{P}_{\zeta }(k) C_{\Lambda}^2T^2(k)  \tilde{W}_k^2(kR(M)) \\
&\ ~~~\equiv C_{\Lambda}^2~\mathcal{A}_{\m{mat,0}}(M),
\label{eq:sigma_P}
}
where $\Omega_{\m{m,0}}$ is the present matter density parameter, $\tilde{W}_k(x)$ is the Fourier function of the window function, $R(M)$ is the comoving scale enclosed the mass $M$ in the background matter density, $\rho_{\rm m,0}$, $T(k)$ is the transfer function for the matter density fluctuations during matter dominated epoch, and
$C_{\Lambda}\approx 0.79$ is the correction for the growth rate during the late-time $\Lambda$ dominated epoch which is estimated by the growth factor in the epoch~\cite{Peebles_PPC}.
In the second line of Eq.~\eqref{eq:sigma_P}, we defined the new mass variance parameter, $\mathcal{A}_{\m{mat,0}}$.
With a use of this parameter, one can estimate the mass variance at an alternative redshift in the matter-dominated epoch by $\mathcal{A}_{\m{mat,0}}/(1+z)^2$.
Hereafter, we will use $\mathcal{A}_{\m{mat,0}}$ as a parameter about the present mass variance.
The expression of $T(k)$ is written by\cite{2008cosm.book.....W}
\begin{align}
T(k) = \cfrac{45}{2} \cfrac{\Omega_\mathrm{m,0}^2 H_0^2}{\Omega_\mathrm{r,0} k^2} \left(-\cfrac{7}{2}+\gamma_\mathrm{E} + \ln{\left(\cfrac{4\sqrt{\Omega_\mathrm{r,0}}k}{\sqrt{3}\Omega_\mathrm{m,0} H_0}\right)} \right)~,
\label{eq:transfer}
\end{align}
where $\Omega_{\m{r,0}}$ is the radiation density parameter at present, and $\gamma_\mathrm{E} \simeq 0.577$ shows the Euler-Mascheroni constant.
This transfer function is valid for the scale, $k\gg 10^{-2} \m{Mpc}^{-1}$. 
These scales are corresponding to the range, $M_{\m{i}}< 10^{15}M_{\odot}$ in terms of the UCMH initial mass. Since we are interested in the mass range which is much smaller than $M_{\m{i}}=10^{15}M_{\odot}$ as we will mention in Sec.~\ref{sec: mcmc}, this function is valid throughout this work.

\begin{figure}[htbp]
    \centering
    \includegraphics[width=8cm,clip]{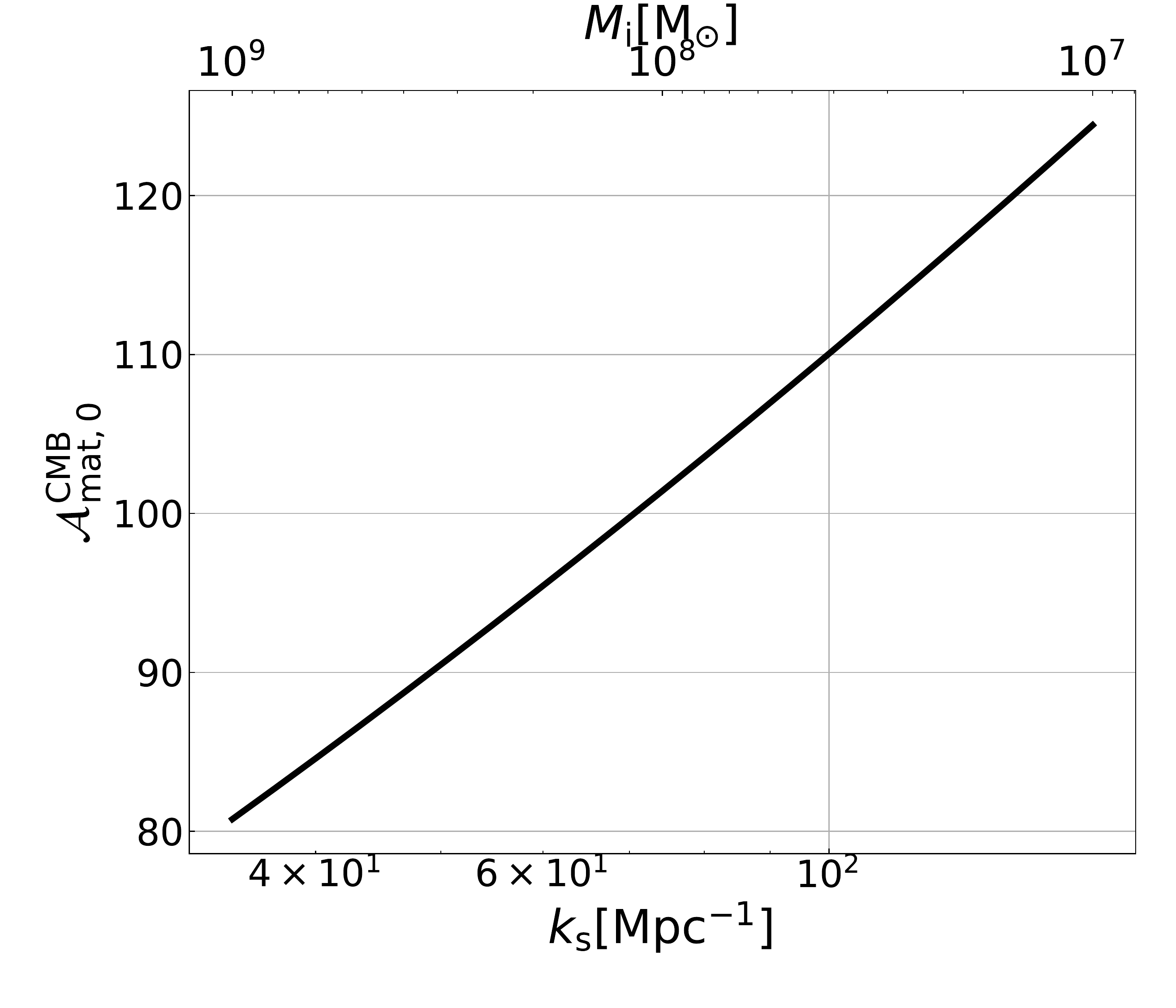}
    \caption{Present mass variance parameter produced by the almost scale-invariant spectrum with the amplitude $\mathcal{A}_{\zeta}^{\m{CMB}}\simeq 2\times 10^{-9}$ in the scale of $k_{\m{s}}$ which corresponds to the initial mass of UCMHs of $M_{\m{i}}$. To make the approximation in Eq.~\eqref{eq: sigma0_M_P_zeta_approx} valid, the value of the mass variance produced by $\mathcal{P}_{\zeta}^{\m{add}}$ at $k_\m{s}$ should be larger than $\mathcal{A}_{\m{mat,0}}^{\m{CMB}}(k_\m{s})$.}
    \label{fig: massvariance_SIPS}
\end{figure}

As we are interested in the mass variance $\mathcal{A}_{\m{mat,0}}(M_{\m{i}})$ where the corresponding scale is $k_{\m{s}}$, the dominant contribution in Eq.~\eqref{eq:sigma_P} would come from the additional power spectrum, $P_{\zeta}^{\m{add}}$(k).
Employing the pointwise window function for $\tilde{W}_{k}$ following the BBKS,
$\mathcal{P}_{\zeta }(k) \tilde{W}_k^2(kR)$ in Eq.~\eqref{eq:sigma_P} has the maximum value at $k_{\m{s}}$. Then Eq.~\eqref{eq:sigma_P} can be approximately written by
\eq{\label{eq: sigma0_M_P_zeta_approx}
\mathcal{A}_{\m{mat,0}}(M_{\m{i}}) \approx  \left. \cfrac{4}{25}~\cfrac{k^4}{\Omega_\mathrm{m,0}^2 H_0^4}~ \mathcal{P}_{\zeta }^{\m{add}}(k) T^2(k) \right|_{k= k_{\rm s}}.
}
Through Eq.~\eqref{eq: sigma0_M_P_zeta_approx}, one can find that the abundance of UCMHs is related to the additional spike-type spectrum properties,~${\mathcal A}_{\zeta}^{\m{add}}$ and $k_{\rm s}$.
This approximation is valid only when the mass variance from the additional spike-type spectrum is much larger than the one from the almost scale-invariant spectrum with the amplitude $\mathcal{A}_{\zeta}^{\m{CMB}}\simeq 2\times 10^{-9}$. 
Figure~\ref{fig: massvariance_SIPS} shows the present mass variance parameter estimated by the almost scale-invariant spectrum. We describe this mass variance parameter as $\mathcal{A}_{\m{mat,0}}^{\m{CMB}}$ to distinguish it from the one estimated in Eq.~\eqref{eq: sigma0_M_P_zeta_approx} hereafter. Therefore, at least the value of the mass variance parameter from the additional spike-type spectrum should be larger than $\mathcal{A}_{\m{mat,0}}^{\m{CMB}}$ represented in Fig.~\ref{fig: massvariance_SIPS} for each initial UCMH mass. We will mention this again in Sec.~\ref{sec: mcmc}.

Before finishing this section, we focus on the baryon gas within UCMHs. 
Of course, to form stars in UCMHs, UCMHs need to enclose baryon gas. Basically, as long as the spike-wave number $k_{\rm s}$ is smaller than the Jeans wave number,~$k_{\rm J}$ at their formation redshift $z_{\m{f}}$, baryon gas can collapse toward the dark matter gravitational potential against their gas pressure.
In that case, we assume that UCMHs would enclose enough baryon gas to form stars with the same mass ratio to dark matter as one of the cosmological backgrounds, $\Omega_{\rm b,0}/\Omega_{\rm m,0}$,
where $\Omega_{\m{b,0}}$ is the baryon density parameter at present although there are other conditions to be satisfied to host stars.
Regarding these conditions, we will mention them in the next section. 
We also assume that the mass ratio between baryon gas and dark matter in UCMHs does not change through their mass evolution represented by Eq.~\eqref{eq: mass_evo_ucmh}.

In addition, we mention the case of that $k_{\rm s}$ is larger than $k_{\rm J}$. 
In that case, although the baryon density fluctuations with $k_{\m{s}}$ cannot evolve due to its own pressure, UCMHs might host the dense baryon gas through the accretion of baryon gas such as the Bondi accretion~\cite{1952MNRAS.112..195B}.
However, the amount of the accreting baryon gas would be small as Ref.~\cite{2020MNRAS.494.4334F} showed. 
Furthermore, since the other conditions would produce stronger constraints on the mass of UCMHs, we neglect this accretion effect.

\section{Pop.~III stars formation in UCMHs}
In high redshifts, $z\lesssim 1000$, the Universe is almost entirely composed of hydrogen (and a few $\%$ helium). In that case, the star would be formed by the zero-metallicity gas, which is called the Pop.~III stars formation. 
UCMHs are minihalos formed at a higher redshift than the standard halo formation scenario. 
Then, if UCMHs fulfill the certain condition to form stars, Pop.~III stars would be formed in UCMHs. 
In this section, we focus on the modification of the global ionization fraction evolution by ionizing photons coming from Pop.~III stars within UCMHs.  

Ionizing photons from Pop.~III stars within UCMHs would create ionized bubbles and evolve them.
The time evolution of the global ionization fraction induced by the ionized bubbles is given by~\cite{2016MNRAS.460..417S}
\eq{\label{eq: xi_evo}
\frac{d x_{\mathrm{e}}}{d t}=\frac{d\left(\zeta f_{\mathrm{coll}}\right)}{d t}-\bar{n}_{H}(t) \alpha_{\mathrm{B}}\left(T_{\m{e}}\right) C_{\mathrm{HII}} x_{\mathrm{e}}
}
where $\bar{n}_{\rm{H}}$, $\alpha_{\rm{B}}$, $T_{\m{e}}$, $C_{\mathrm{HII}}$, and $f_{\m{coll}}$ are the mean number density of hydrogen nucleus in the IGM, the case B recombination rate given in Ref.~\cite{Fukugita&Kawasaki_cooling}, the electron temperature, the clumping factor, and the collapsed fraction respectively.  
We set the electron temperature as $T_{\m{e}}=10^4\m{K}$, and the clumping factor as $C_{\mathrm{HII}}\approx 3$ in this work. 
The star ionizing efficiency $\zeta$ is decomposed by $\zeta \equiv A_{\mathrm{He}} N_{\gamma} f_{\mathrm{esc}}f_{\star}$, where $A_{H_{\m{e}}}\approx 1.22$ is the correction factor for singly ionized helium, $N_{\gamma}$ shows the average number of ionizing photons produced per stellar baryon, $f_{\mathrm{esc}}$ represents the escape fraction of ionizing photons, and $f_{\star}$ is the the star formation efficiency.
Since contributions of Pop.~III stars to the ionization history is mostly unconstrained at present~\cite{2017ApJ...843..129B}, we adopt a toy model for their effect mimicking the approach for Pop.~II stars of the previous work~\cite{2020PhRvD.101l3518I}. 
We set $N_{\gamma}=4\times 10^4$ which is anticipated for the hotter
photospheres of these metal-free stars~\cite{2001ApJ...552..464B}. 
For the escape fraction, we assume that all ionizing photons escape expected for star formations in small halos, so that $f_{\m{esc}}=1$. 
We also set $f_{\ast}=5\times 10^{-4}$ which is on the lower end of the value typically used in Refs.~\cite{2009ApJ...694..879T,2018MNRAS.475.5246V}, although the most commonly used value is about $10^{-3}$.
Note that the value of the escape fraction can be taken in one of the star formation efficiency.

With the common assumption that the star formation rate in a halo is proportional to the baryon gas accretion rate $\dot{M}_{\m{acc}}$ with $f_{\star}$ as the proportional coefficient, the first term in right-hand side of Eq.~\eqref{eq: xi_evo} can be written
\eq{\label{eq: ionphoton_ucmh}
\frac{\mathrm{d}\left(\zeta f_{\mathrm{coll}}\right)}{\mathrm{d} t}&= A_{\mathrm{He}} N_{\gamma} f_{\mathrm{esc}} \int_{M_{\m{min}}}\hspace{-3mm}\mathrm{d}M^\prime~f_{\star} \dot{M}_{\mathrm{acc}}\frac{dn(M^{\prime}, z)}{dM}\frac{1}{\bar{\rho}_{b}(z)}\\
&\ =A_{\mathrm{He}} N_{\gamma} f_{\mathrm{esc}}\frac{\dot{\rho}_{\m{SFR}}(z)}{\bar{\rho}_{\m{b}}(z)},
}
where $\dot{\rho}_{\m{SFR}}$ is the star formation rate density, $\bar{\rho}_{\m{b}}$ is the mean baryon density, $\dot{M}_{\mathrm{acc}}$ can be estimated from Eq.~\eqref{eq: mass_evo_ucmh} as
\eq{\label{eq: mass_acc_ucmh}
\dot{M}_{\mathrm{acc}}(M, z) = M_{\m{i}}H(z)\frac{\Omega_{\rm b,0}}{\Omega_{\rm m,0}},
}
and $dn/dM$ is the UCMH mass function associated with Eq.~\eqref{eq: num_dens_ucmh}. In Eq,~\eqref{eq: ionphoton_ucmh}, $M_{\m{min}}$ represents the criteria of the halo mass to form the Pop.~III stars including the Lyman-Werner negative feedback~\cite{2001ApJ...548..509M,2007ApJ...671.1559W},
\eq{\label{eq: min_mass_popIII}
M_{\mathrm{min}}=M_{\m{h}}\left(T_{\mathrm{vir}}=500 \mathrm{~K}\right)\left[1+6.96 F_{\m{L W, 21}}^{0.47}\right],
}
with $M_{\m{h}}(T_{\mathrm{vir}})$ which is the typical halo mass with a given virial temperature $T_{\mathrm{vir}}$, and the Lyman-Werner intensity integrated over a solid angle, $F_{\m{L W, 21}}$ which is in units of $10^{-21} \m{erg} \m{s^{-1}}\m{Hz^{-1}} \m{cm^{-2}}$. 

To estimate $M_{\m{h}}(T_{\mathrm{vir}}=500 \mathrm{~K})$, we use the relation,
\eq{\label{eq: Mvir_temp}
M&_\m{h}(T_{\m{vir }},z_\m{f})\\
&\ \approx 4\times 10^{5}~\mathrm{M}_{\odot}\lr{\frac{T_{\m{vir}}}{500\m{K}}\cfrac{1+z_\m{f}}{10}}^{\frac{3}{2}}\left(\cfrac{\Omega_\m{m,0}}{\Omega_\m{m}(z_\m{f})}\right)^{-\frac{1}{2}},
} 
where $\Omega_\m{m}(z)$ is the matter density parameter
at a redshift $z$ after the matter-dominant era,
\eq{
\Omega_{\m{m}}(z)=\frac{\Omega_{\m{m,0}}(1+z)^{3}}
{\Omega_{\m{m,0}}(1+z)^{3}+\Omega_{\Lambda,0}}~,
}
with $\Omega_{\Lambda,0}$ which is the present density parameter of the cosmological constant.

We also employ the relation between the Lyman-Werner intensity and $\dot{\rho}_{\m{SFR}}$~\cite{2015MNRAS.453.4456V,2018MNRAS.479.4544M} and estimate $F_{\m{L W, 21}}$ as
\eq{\label{eq: FLW21}
F_{\m{L W, 21}}=7.22 \frac{(1+z)^{3}}{H(z)} e^{-\tau_{L W}}N_{\mathrm{LW}} \frac{\dot{\rho}_{\mathrm{SFR}}}{\bar{\rho}_{\m{b}}},
}
where $H(z)$ is the Hubble parameter, $\tau_{\m{LW}}$ is the intergalactic opacity for the Lyman-Werner photons, and $N_{\m{LW}}$ is the number of Lyman-Werner photons produced per baryon in stars. 
Here we set $e^{-\tau_{\m{L W}}}=0.5$~\cite{2001ApJ...560..580R,2007ApJ...665...85J} while the value of $\tau_{\m{LW}}$ might increase due
to UCMHs obscuring the Lyman-Werner background.
We also set $N_{\mathrm{LW}}=10^5$ for Pop.~III stars~\cite{2018MNRAS.479.4544M}.

In Eq.~\eqref{eq: min_mass_popIII}, we assume that $T_{\m{vir}}=500\m{K}$ is hot enough to excite rotational transitions of molecular hydrogen, which can be an effective cooling mechanism to form Pop.~III stars. In this work, we assume that UCMHs can host Pop.~III stars as long as their mass is larger than this minimum mass, $M_{\m{min}}$, and the Jeans mass. It is noted that the value of $M_{\m{min}}$ is larger than the Jeans mass in our interesting parameter ranges.


UCMHs would have the mass variety even in the case of delta function type power spectrum of Eq.~\eqref{eq: add_spike_pk} because there is a variety of their formation redshifts $z_{\m{f}}$ following the number density formed in the duration of [$z_{\m{f}},z_{\m{f}}+\Delta z_{\m{f}}]$,
\begin{equation}
\frac{dn}{dz_{\rm f}} =\frac{k_{\rm s}^3}{1+z_{\rm f}}
h\lr{\nu=\frac{\delta_{\rm c}(1+z_{\m{f}})}{\mathcal{A}_{\m{mat,0}}^{1/2}}}.
\label{eq: dndzf_ucmh}
\end{equation}
In Eq.~\eqref{eq: dndzf_ucmh}, $h(\nu)$ is given by
\begin{equation}
h(\nu)=\frac{\nu}{(2\pi)^2 3^{3/2}}e^{-\nu^2/2}f(\nu).
\label{hnu}
\end{equation}
The earlier formed UCMHs are heavier than the later formed ones due to matter accretion. 
However, in peak theory, most UCMHs are formed around the specific redshift $z_{\m{eff}} \approx 2.1936\mathcal{A}_{\m{mat,0}}^{1/2}/\delta_{c}-1$ which can be estimated by the derivative of the differential number density of Eq.~\eqref{eq: dndzf_ucmh} unlike the Press Schechter halo formation formalism~\cite{1974ApJ...187..425P}.
Then we assume that the mass of all UCMHs is given by
\eq{\label{eq: effective_mass_ucmh}
M_{\m{eff}}(z) \equiv M(z,z_{\m{eff}})  = M_{\m{i}}~\m{Max}\lr{1,1+\ln  \left( \frac{1+z_{\m{eff}}}{1+z} \right)},
}
and the mass function of UCMHs is given by
\eq{
\frac{dn(M,z)}{dM}\approx n(M,z)\delta(M-M_{\m{eff}}).
}
Therefore, we approximately calculate Eq.~\eqref{eq: ionphoton_ucmh} as
\eq{
\frac{\mathrm{d}\left(\zeta f_{\mathrm{coll}}\right)}{\mathrm{d} t}\approx
\begin{cases}
A_{\mathrm{He}} f_{\gamma} f_{\mathrm{esc}} f_{\star}\dot{M}_{\mathrm{acc}}n(M_{\m{eff}},z) &  M_{\m{eff}}>M_{\m{min}} \\
0 &\ \textit{otherwise},
\end{cases}
}

\begin{figure}[htbp]
    \centering
    \includegraphics[width=8cm,clip]{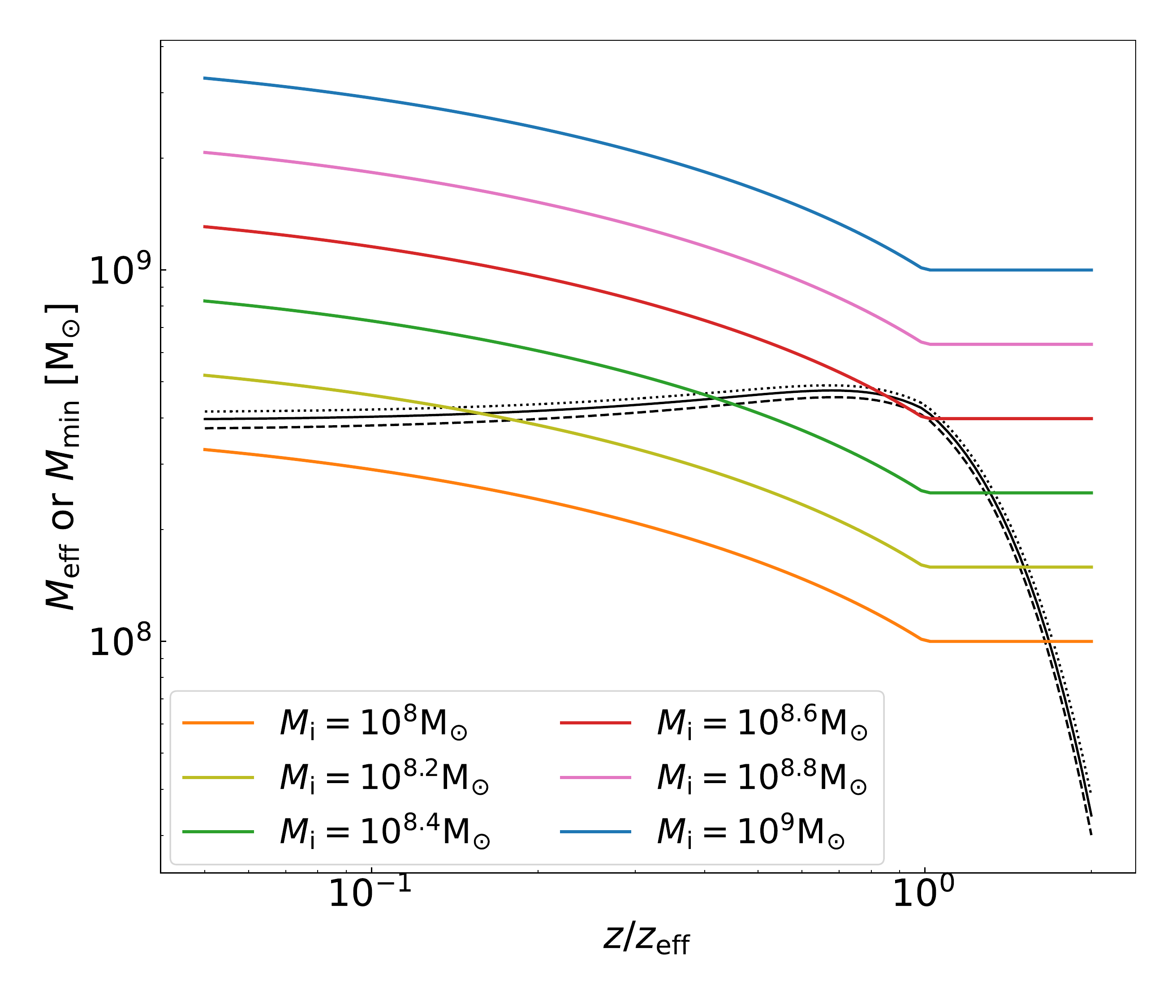}
    \caption{The minimum mass of UCMHs to host Pop.~III stars in Eq.~\eqref{eq: min_mass_popIII} (black line) and the UCMH effective mass represented in Eq.~\eqref{eq: effective_mass_ucmh}. The horizontal axis shows the redshift normalized by the $z_{\m{eff}}$ with $\mathcal{A}_{\m{mat,0}}$. The vertical axis shows the minimum and the effective mass of UCMHs in a unit of the solar mass. The black dotted, solid and dashed lines represent the minimum masses with $\mathcal{A}_{\m{mat,0}}=100,200,500,$ respectively. The colorful solid lines show the effective masses with the different initial mass models.}
    \label{fig: mass_evo}
\end{figure}

Figure.~\ref{fig: mass_evo} shows $M_{\m{eff}}$ and $M_{\m{min}}$ evolution in terms of the redshift.
It is noted that there is no dependency of the minimum mass $M_{\m{min}}$~(so that $\dot{\rho}_{\m{SFR}}$) in Eq.~\eqref{eq: min_mass_popIII} on the UCMH intial mass. This is because the dependencies on the UCMH mass for both $\dot{M}_{\m{acc}}$ and $dn/dM$ in Eq.~\eqref{eq: ionphoton_ucmh} cancel each other. In addition, the dependency on $\mathcal{A}_{\m{mat,0}}$ is also weak because the redshift dependence of the bracket term in Eq.~\eqref{eq: min_mass_popIII} and the other term, $M_{\m{h}}(T_{\m{vir}}=500)$ are almost canceled out.

\section{Reionization model including UCMH Pop.~III stars}\label{sec: reio_model}
In this work, we consider the global ionization history adding the effects of UCMH Pop.~III stars on the standard ionization scenario, where the Universe will be reionized much after the recombination epoch. 
In our scenario, the effects of UCMH Pop.~III stars on the cosmic reionization could be dominant only at high redshifts, $z\gtrsim 10$, and after that, Pop.~II stars and first galaxies become the main ionizing photon sources.
Therefore, taking into account the ionizing photons from UCMH Pop.~III stars for the cosmic reionization, we assume that the evolution of the global ionization fraction can be decomposed into three terms,
\eq{\label{eq: ion_model}
x_{\m{e}}(z)=&(x_{\m{e}}^{\m{rec}}(z)+x_{\m{e}}^{\m{reio}}(z))\\
&\ ~~~~+x_{\m{e}}^{\m{add}}(z)(1-x_{\m{e}}^{\m{rec}}(z)-x_{\m{e}}^{\m{reio}}(z)),
}
where $x_{\m{e}}^{\m{rec}}$ is the global ionization fraction in the recombination epoch, and $x_{\m{e}}^{\m{reio}}$ represents the contribution from the main reionization source including Pop II stars and galaxies.
For obtaining $x_{\m{e}}^{\m{rec}}$, we employ the recombination code~{\tt RECFAST}~\cite{1999ApJ...523L...1S,2000ApJS..128..407S,2008MNRAS.386.1023W,2009MNRAS.397..445S},
whereas for $x_{\m{e}}^{\m{reio}}$, we adopt the widely used ``tanh" model as~\cite{2008PhRvD..78b3002L}
\begin{eqnarray}
 x_{\m{e}}^{\m{reio}}(z)&=&x_{\m{e}}^{\mathrm{before}}+\frac{1}{2}\left(x_{\m{e}}^{\m{after}}-x_{\m{e}}^{\mathrm{before}}\right)
 \nonumber \\
&&\quad \
 \times \left[1+\tanh \left(\frac{y^{\mathrm{reio}}-y(z)}{\Delta y}\right)\right],
 \label{eq:tanh-shape}
\\
y(z)&=&(1+z)^{3 / 2},
\end{eqnarray}
where
$y^{\m{reio}}=y(z^{\m{reio}})$, $\Delta y=1.5 \sqrt{1+z^{\m{reio}}} \Delta z$ with the duration of reionizaiton, and we set $\Delta z=0.5$.
Note that $x_{\m{e}}^{\m{reio}}(z^{\m{reio}})=0.5$.
In Eq.~\eqref{eq:tanh-shape},
$x_{\mathrm{e}}^{\mathrm{after}}=1$ is the ionization fraction after finishing reionization, and $x_{\mathrm{e}}^{\mathrm{before}}$ is the leftover ionization fraction well after the recombination epoch adopted as $x_{\mathrm{e}}^{\mathrm{before}}\sim 10^{-4}$. 
The evolution of $x_{\m{e}}^{\m{reio}}$(z) is characterized only by the value of $z_{\m{reio}}$. 
We deal with $z_{\m{reio}}$ as a free parameter in the following analysis. 

For the additional ionization fraction $x_{\m{e}}^{\m{add}}$, we estimate by solving Eq.~\eqref{eq: xi_evo}.
Generally, one cannot add ionization fractions from several sources. To calculate the time evolution of the global ionization fraction properly, one needs to sum up the photoionization rates and solve the balance between recombination and the ionization rate. 
However in Eq.~\eqref{eq: xi_evo}, one calculates the contribution of the ionized bubbles surrounding Pop.III stars on the global ionization fraction while the terms $x_{\m{e}}^{\m{rec}}$ and $x_{\m{e}}^{\m{reio}}$ are related to global ionization fraction in the IGM.
Therefore, we here assume that the global ionization fraction can be estimated by Eq.~\eqref{eq: ion_model} in our model.
Although Eq.~\eqref{eq: ion_model} is written with respect to redshift, Eq.~\eqref{eq: xi_evo} is written with respect to cosmic time.
Therefore, to estimate three types of the ionization fraction in Eq.~\eqref{eq: ion_model} consistently, we need to be careful about the time (redshift) step when calculating Eqs.~\eqref{eq: xi_evo} and~\eqref{eq: ion_model}. We put the lower limit of the redshift, $z_{\m{cut}}$, when calculating $x_{\m{e}}^{\m{add}}$. This is because we would like to focus on the contribution coming from halos created by the additional spike-type power spectrum rather than the standard halos originating from  the almost scale-invariant spectrum.
Therefore, we set $z_{\rm cut}$ to the redshift at which the standard halos are formed effectively, so that
$z_{\m{cut}}=\mathcal{A}_{\m{mat,0}}^{\m{CMB}}/\delta_ c-1$.

\begin{figure}[htbp]
    \centering
    \includegraphics[width=8cm,clip]{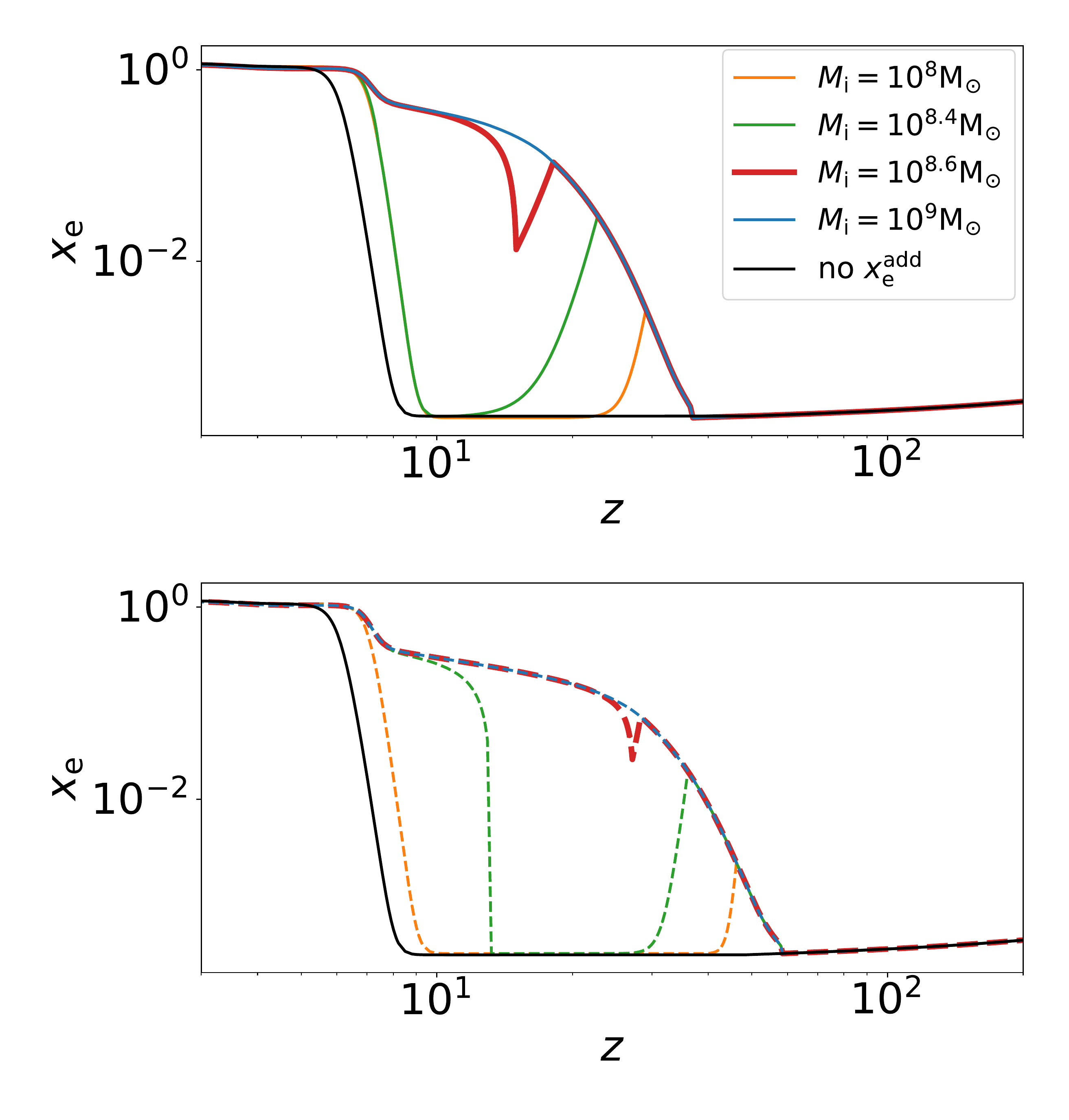}
    \caption{The cosmological evolution of the global ionization fraction estimated by Eq.~\eqref{eq: ion_model}. The color difference shows the difference of the initial mass of UCMHs, and the different line style shows the different value of $\mathcal{A}_{\m{mat,0}}$: $\mathcal{A}_{\m{mat,0}} = 200$~(solid) and $\mathcal{A}_{\m{mat,0}}=500$~(dashed). The black solid line shows the global ionization fraction evolution without any additional contribution from Pop.~III stars in UCMHs so that $x_{\m{e}}^{\m{add}}=0$. In this figure, we set $z_{\m{reio}}=6$. The extent of the contributions from UCMHs with a larger initial mass than $M_{\m{i}}=10^{8.6}\m{M}_{\odot}$ would saturate. Thus, the lines for $M_{\m{i}}=10^{8.6}\m{M}_{\odot}$ and $M_{\m{i}}=10^{9}\m{M}_{\odot}$ mostly overlap. }
    \label{fig: ion_frac_evo}
\end{figure}

Figure.~\ref{fig: ion_frac_evo} shows the examples of the ionization fraction evolution obtained from Eq.~\eqref{eq: ion_model}. 
In this figure, the color difference shows the difference in the initial mass of UCMHs, and the different line style shows the different value of $\mathcal{A}_{\m{mat,0}}$. The black solid line shows the global ionization fraction evolution without any additional contribution from Pop.~III stars in UCMHs.
In this figure, we set $z_{\m{reio}}=6$. 
We found that roughly speaking, the initial mass of UCMH, $M_{\m{i}}$, controls the extent of the ionizing effect, and the mass variance on the scale of $k_{\m{s}}$, $\mathcal{A}_{\m{mat,0}}$, determines the redshift where the effect begins to work.
The nonmonotonic behavior comes from the duration when UCMHs host Pop.~III stars.
As one can see in Fig.~\ref{fig: mass_evo}, if the UCMH initial mass is in the range $10^{8}\m{M}_{\odot} \lesssim M_{\m{i}}\lesssim 10^{8.6}\m{M}_{\odot}$, UCMHs would host Pop.~III stars in two separate periods. 
Then the global ionization through UCMHs would occur in three phases: (1) increments of $x_{\m{e}}$ by ionizing photons from Pop.~III stars hosted by UCMHs which are in the first period, (2) recombination due to no Pop.~III stars in UCMHs, and (3) increments of $x_{\m{e}}$ again by UCMHs which are in the second period.
It is also noted that the extent of the contributions from UCMHs with a larger initial mass than $M_{\m{i}}=10^{8.6}\m{M}_{\odot}$ would saturate. Thus, in Fig.~\ref{fig: ion_frac_evo}, the lines for $M_{\m{i}}=10^{8.6}\m{M}_{\odot}$ and $M_{\m{i}}=10^{9}\m{M}_{\odot}$ overlap.
We also mention the contributions from UCMHs with a smaller initial mass than $10^{8}\m{M}_{\odot}$ in that case, the contributions are too small to change the global ionization evolution.

\begin{figure}[htbp]
    \centering
    \includegraphics[width=8cm,clip]{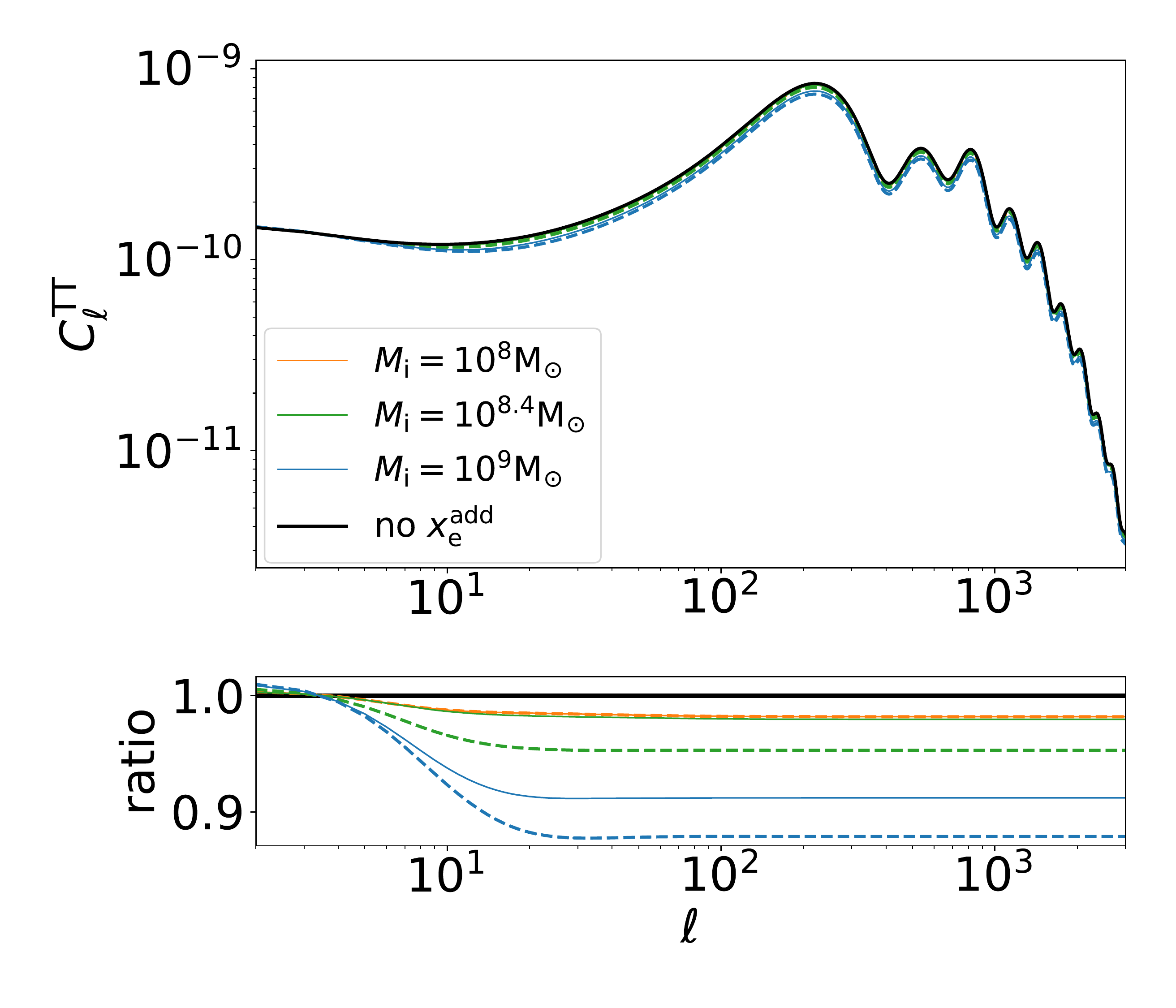}
    \caption{[Top panel]: Angular power spectrum of the CMB temperature. The colorful lines shows CMB temperature anisotropies with the three different initial mass of UCMHs, and the different line style shows the different value of $\mathcal{A}_{\m{mat,0}}$: $\mathcal{A}_{\m{mat,0}} = 200$~(solid) and $\mathcal{A}_{\m{mat,0}}=500$~(dashed). The black solid line shows the standard CMB temperature anisotropy without any additional effects from Pop.~III stars in UCMHs, $x_{\m{e}}^{\m{add}}=0$. [Bottom panel]: Ratio between the standard CMB temperature anisotropy (black solid line) and one of each UCMH initial mass model.}
    \label{fig: cell_tt_popIII}
\end{figure}

\begin{figure}[htbp]
    \centering
    \includegraphics[width=8cm,clip]{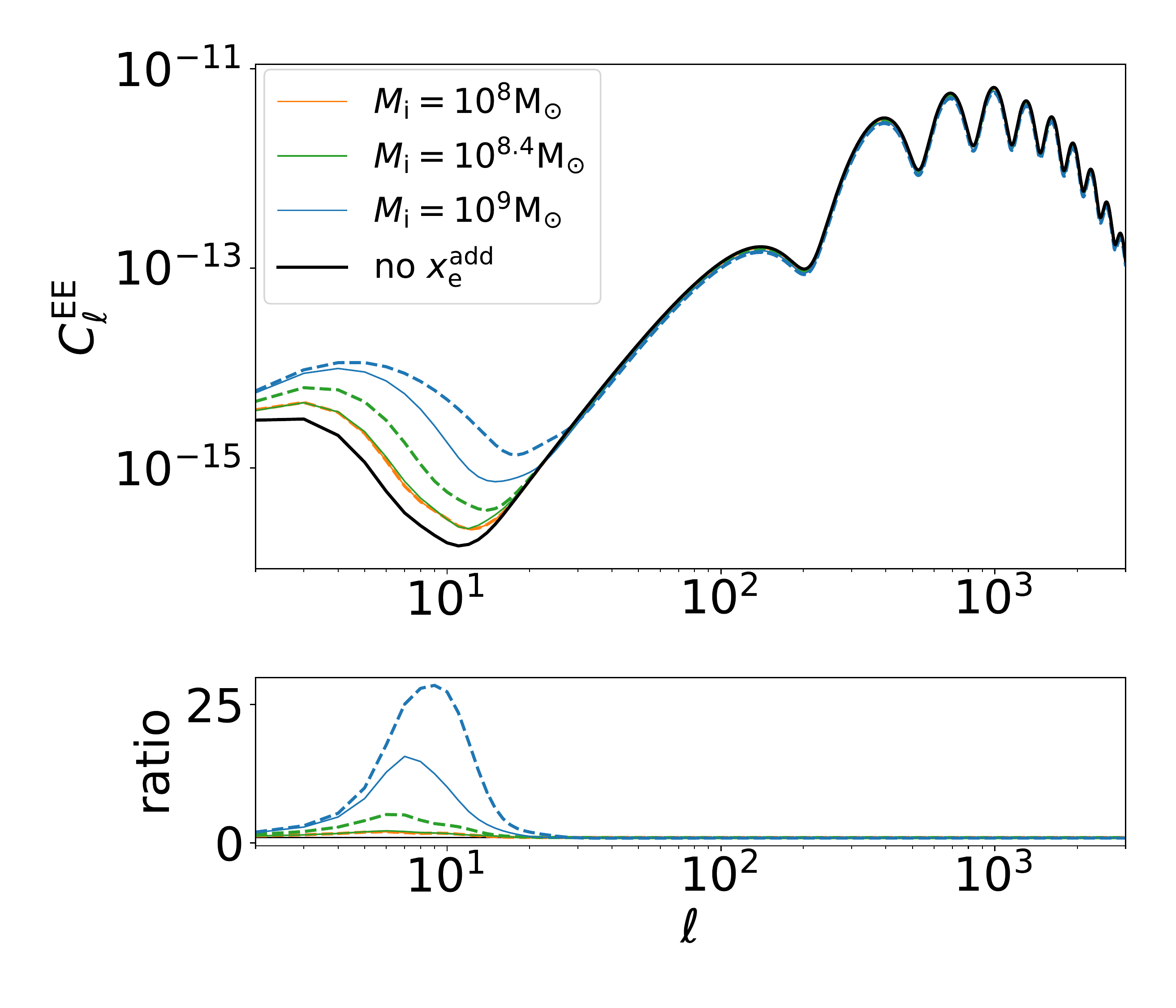}
    \caption{[Top panel]: Angular power spectrum of the CMB E-mode polarization. The colorful lines shows CMB E-mode polarization anisotropies with the three different initial mass of UCMHs, and the different line style shows the different value of $\mathcal{A}_{\m{mat,0}}$: $\mathcal{A}_{\m{mat,0}} = 200$~(solid) and $\mathcal{A}_{\m{mat,0}}=500$~(dashed). The black solid line shows the standard CMB E-mode polarization anisotropy without any additional effects from Pop.~III stars in UCMHs, $x_{\m{e}}^{\m{add}}=0$. [Bottom panel]: Ratio between the standard CMB E-mode polarization anisotropy (black solid line) and one of each UCMH initial mass model.}
    \label{fig: cell_ee_popIII}
\end{figure}

One of the useful observables to test these examples would be the CMB anisotropy.
Figures~\ref{fig: cell_tt_popIII} and~\ref{fig: cell_ee_popIII} show the temperature and the E-mode polarization anisotropies for different UCMH initial mass models.
A good indicator to explain these modifications in Figs.~\ref{fig: cell_tt_popIII} and~\ref{fig: cell_ee_popIII} is the Thomson scattering optical depth for CMB photons which is defined by, 
\eq{\label{eq: optical_depth}
\tau = \int \frac{d z}{H(z)} \sigma_{\rm T } x_{\m{e}}(z) \bar{n}_{H}(z),
}
where $\sigma_{\m{T}}$ is the Thomson cross section.
As one can see in Fig.~\ref{fig: cell_tt_popIII}, the amplitude of the temperature anisotropy is suppressed as the optical depth increase. The theoretical dependency is $\propto \exp(-\tau)$, and the suppression is slight in these models.
For the E-mode polarization anisotropy, one can find a significant difference among these models in Fig.~\ref{fig: cell_ee_popIII}, especially with the behavior at $\ell\lesssim 20$ mode. This behavior is called the ``reionization bump" whose amplitude is proportional to $\tau^2$.

\section{MCMC analysis with Planck 2018}\label{sec: mcmc}
\subsection{Setup}\label{sec:MCMCsetup}
In order to constrain or investigate the effect of ionizing photons from UCMH Pop III stars based on our model in Eq.~\eqref{eq: ion_model}, we employ the MCMC analysis with Planck 2018 data.
Chains of MCMC samples are generated by the publicly open code {\tt MontePython}~\cite{Audren:2012wb}, which adopts the code {\tt CLASS}~\cite{Blas_2011} for calculating the theoretical CMB angular power spectrum.
We have modified the {\tt CLASS} code including the global ionizing effect originating from Pop.~III stars in UCMHs represented in Eq.~\eqref{eq: ion_model}.
In the calculation, we deal with the following two parameters as free parameters: (1)the mass variance on the scale $k_{\m{s}}$ represented by $\mathcal{A}_{\m{mat,0}}$ which relates to the amplitude of the spike power spectrum, and (2)the standard reionization parameter $z_{\m{reio}}$ in Eq.~\eqref{eq:tanh-shape}. 
For the initial mass of UCMHs $M_{\m{i}}$ which is corresponding to the spike scale $k_s$, we fix to some value in range of $M_{\m{i}}=10^{7-9}M_{\odot}$ for each calculations. 
We set a lower limit for $\mathcal{A}_{\m{mat,0}}$ to make the approximation represented in Eq.~\eqref{eq: sigma0_M_P_zeta_approx} valid.
That is, we put hard priors for $\mathcal{A}_{\m{mat,0}}$ to satisfy the condition where $\mathcal{A}_{\m{mat,0}}(M_{i})$ is larger than $\mathcal{A}_{\m{mat,0}}^{\m{CMB}}(M_{i})$ shown in Fig.~\ref{fig: massvariance_SIPS} for each $M_{\m{i}}$. 
For example, we set $\mathcal{A}_{\m{mat,0}}(10^8\m{M}_{\odot})>79.2$ and $\mathcal{A}_{\m{mat,0}}(10^9\m{M}_{\odot})>64.7$.
We also assume that the standard "tanh"-type reionization occurs by $z=3$, so that we set $z_{\m{reio}}>3$.

We should mention the other cosmological parameters. 
We are primarily interested in the global ionization fraction evolution in Eq.~\eqref{eq: xi_evo} and the resultant optical depth that appeared in the reionization bump. The three parameters ($M_{\m{i}}$, $\mathcal{A}_{\m{mat,0}}$, $z_{\m{reio}}$) mainly control them.
Therefore, we fix other cosmological parameters to the Planck best-fit parameter of the TT, TE, EE, low-$\ell$ + lensing measurement, $\Omega_{\m{b,0}}=0.02237$, $\Omega_{\m{cdm,0}}=0.1200$, $100\theta_{\m{s}}=1.04092$, $\m{ln}10^{10}\mathcal{A}_{\zeta}^{\m{CMB}}=3.044$, and $n_s=0.9649$.
These parameters do not affect the reionization bump much.

It also should be noted about the accuracy of the MCMC analysis. 
In order to obtain accurate results from MCMC methods, it is essential to check whether the MCMC chains contain enough samples which are independent of each other and cover a sufficient volume of parameter space. Otherwise, the density of the samples will not converge to the actual posterior probability distribution. Therefore in this work, we employ the Gelman and Rubin convergence statistic $R$ which represents the ratio of the variance of parameters between chains to the variance within each chain, and run the analysis by the chains that will satisfy the condition, $R-1<0.05$ ~\cite{1992StaSc...7..457G,doi:10.1080/10618600.1998.10474787}.

\subsection{MCMC results and discussion}
\begin{figure*}[htbp]
    \centering
    \includegraphics[width=16cm,clip]{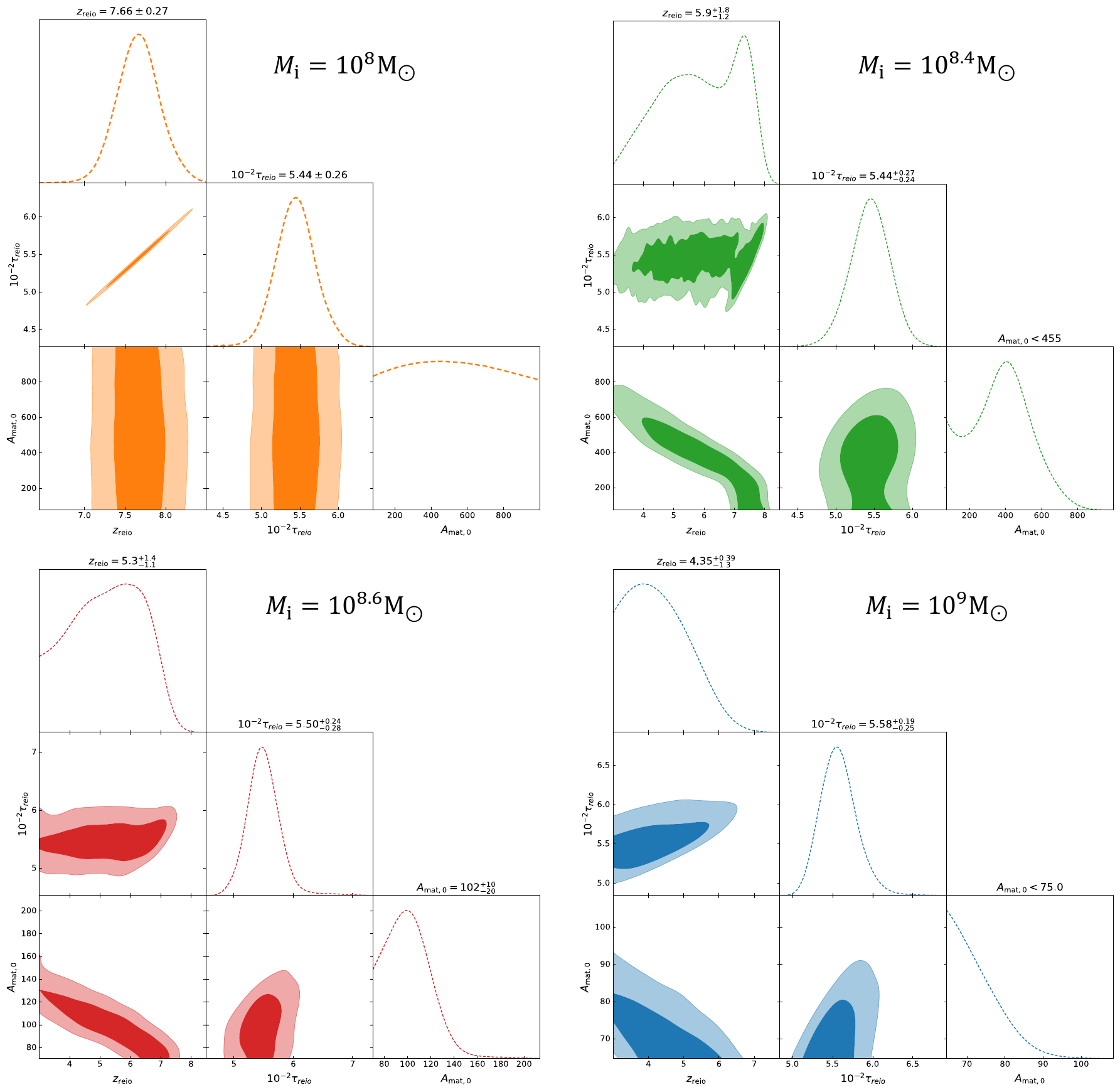}
    \caption{MCMC results for several models of the UCMH initial mass, $M=10^8\m{M}_{\odot}$~(top left), $M=10^{8.2}\m{M}_{\odot}$~(top right), $M=10^{8.4}\m{M}_{\odot}$~(bottom left), $M=10^9\m{M}_{\odot}$~(bottom right). $z_{\m{reio}}$ and $\mathcal{A}_{\m{mat,0}}$ are free paramters, and $\tau$ is the derived paramter. The thick shaded region shows the $1\sigma$ region, and the thin shaded region shows the $2\sigma$ region.}
    \label{fig: mcmc_popIII_sum}
\end{figure*}

We have implemented the MCMC analysis with the UCMH initial mass in the range of $10^{7}M_{\odot}<M_{\m{i}}<10^{9}M_{\odot}$. Figure~\ref{fig: mcmc_popIII_sum} shows the MCMC results for our ionization history model represented by Eq.~\eqref{eq: ion_model} with the four UCMH initial mass values, $M_{\m{i}}=10^8 M_{\odot}$, $10^{8.2} M_{\odot}$, $10^{8.4} M_{\odot}$, $10^{9}M_{\odot}$ as examples. 
In this analysis, $A_{\m{mat,0}}$ and $z_{\m{reio}}$ are free parameters, and the optical depth $\tau$ is derived parameter from Eq.~\eqref{eq: optical_depth} with the sampling data of above two parameters. 
In this figure, the thick colored region shows the $1\sigma$ region, and the thin colored region represents the $2\sigma$ region.
As explained in Sec.~\ref{sec: reio_model}, the CMB polarization anisotropy is sensitive to the optical depth $\tau$ during and after the cosmic reionization, and the Planck measurement basically provides the constraint on $\tau$.
We found that even in our ionization history which includes the effect induced by UCMH Pop. III stars, the Planck observation data (TT, TE, EE, low-$\ell$ + lensing) prefer the almost same best-fit value for the optical depth, $\tau^{\m{CMB}}=0.0544$. 
Therefore, as the additional ionization fraction $x_{\m{e}}^{\m{add}}$ increases the optical depth to some extent, the value of $z_{\m{reio}}$ is shifted to lower than the Planck result, $z_{\m{reio}}^{\m{CMB}}=7.67$.
If the mass of UCMHs is smaller than the lower mass criteria estimated by Eq.~\eqref{eq: min_mass_popIII}, the effect from UCMH Pop.~III stars does not work no matter how large the $\mathcal{A}_{\m{mat,0}}$ is.

On the other hand, the initial mass of UCMHs is larger than the $10^8M_{\odot}$, ionizing photons from UCMH Pop.~III stars produce non-negligible optical depth, and the small $z_{\m{reio}}$ would be preferred to compensate the increment of the optical depth as you can infer from Figs.~\ref{fig: mass_evo} and~\ref{fig: ion_frac_evo}.
If the UCMH initial mass is larger than $10^{8.2}M_{\odot}$, UCMHs could host Pop.~III stars for a long duration. 
Then the contribution would drastically increase, and if the initial mass is larger than $M_{\m{i}}>10^{8.6}M_{\odot}$, the contribution would be maximum.
In that case, the decrement of $z_{\m{reio}}$ can not cancel out the increase of $\tau$ coming from large $\mathcal{A}_{\m{mat,0}}$, and the constraint on the value of $\mathcal{A}_{\m{mat,0}}$ becomes stringent.
In this work, we studied this model for the UCMH initial mass ranges, $M_{\m{i}}<10^9\m{M}_{\odot}$, however, this constraint would be valid for the larger initial mass case, $M_{\m{i}}>10^{9}\m{M}_{\odot}$.

Although this work includes several uncertainties about astrophysics, it might be useful to convert the constraint on the $\mathcal{A}_{\m{mat,0}}$ to the one of $\mathcal{A}_{\zeta}^{\m{add}}$ which shows the amplitude of the additional spike-type power spectrum quantitatively.
Considering the relation between $\mathcal{A}_{\m{mat,0}}$ and $\mathcal{A}_{\zeta}^{\m{add}}$ represented by Eq.~\eqref{eq: sigma0_M_P_zeta_approx}, one can find
\eq{\label{eq: convert_Amat_Azeta}
\mathcal{A}_{\zeta}^{\m{add}}\approx 10^{-9} \mathcal{A}_{\m{mat,0}}\lr{8.7-\frac{1}{3}\ln\lr{\frac{M_{\m{i}}}{10^6\m{M_{\odot}}}}}^{-2}.
}

\begin{figure}[htbp]
    \centering
    \includegraphics[width=8cm,clip]{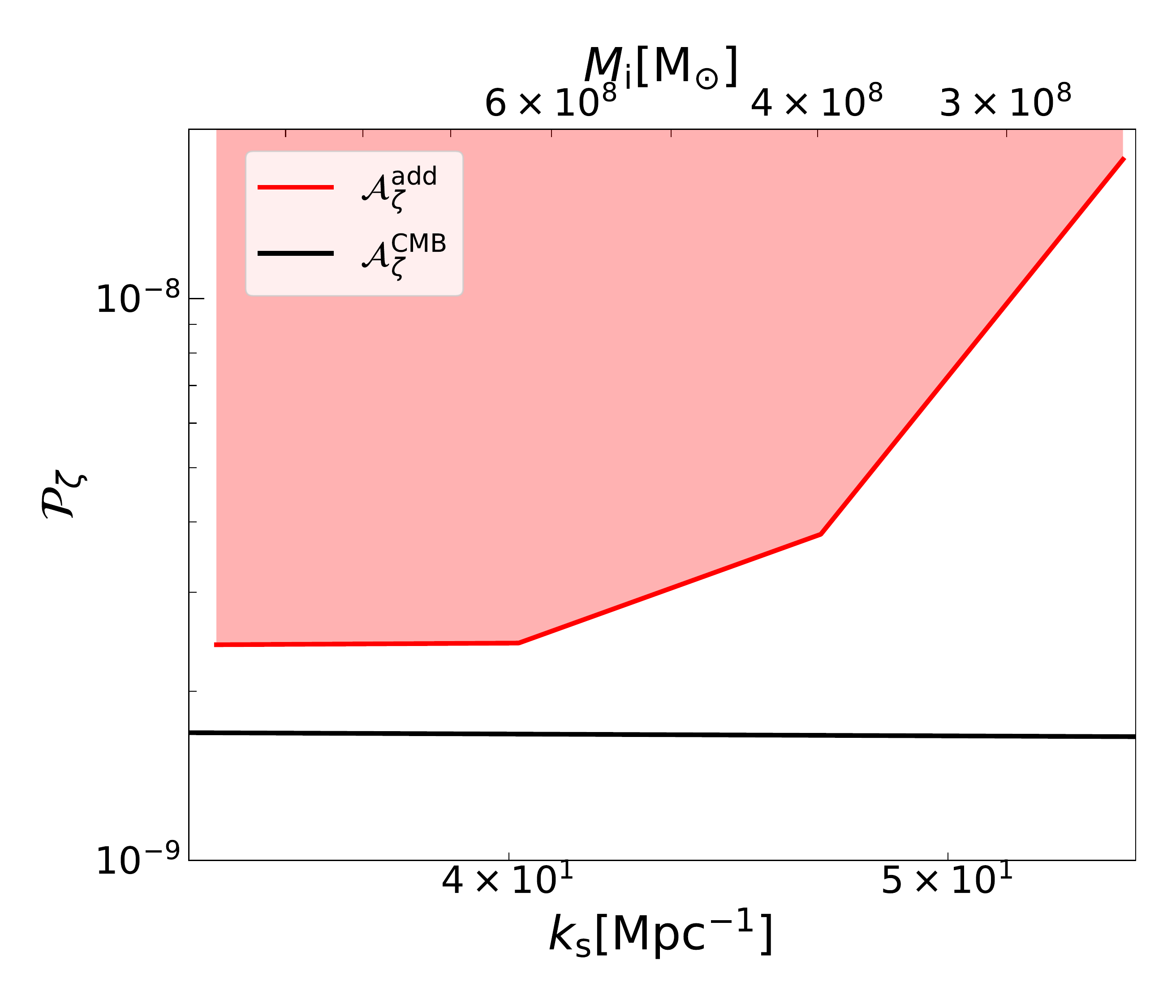}
    \caption{Upper limit on the amplitude of the additional spike-type power spectrum, $\mathcal{A}_{\zeta}^{\m{add}}$, corresponding to the 2-$\sigma$ constraint of the $\mathcal{A}_{\m{mat,0}}$ through the MCMC analysis with Planck 2018 data.
    It is noted that we assume $z_{\m{reio}}>3$ in this figure.
    The red shaded region shows the limited parameter region through this work.
    The solid black line shows the almost scale-invariant power spectrum with the amplitude $\mathcal{A}_{\zeta}^{\m{CMB}}\simeq 2\times 10^{-9}$ with the spectral index, $n_{\m{s}}=0.9649$.}
    \label{fig: Pzeta_const}
\end{figure}

Figure~\ref{fig: Pzeta_const} shows the upper limit of $\mathcal{A}_{\zeta}^{\m{add}}$ by Planck 2018 data.
This limit is estimated through Eq.~\eqref{eq: convert_Amat_Azeta} with the 2-$\sigma$ constraint of the $\mathcal{A}_{\m{mat,0}}$ obtained by the MCMC analysis.
It is noted that in this figure we assume $z_{\m{reio}}>3$ as we set the lower limit in the MCMC analysis.
Since UCMHs with a larger initial mass than $10^{8.4}\m{M}_{\odot}$ are powerful for the global ionization in our model, the large excess of the primordial power spectrum in scales, $k_{\m{s}}\lesssim 40{\m{Mpc}^{-1}}$ is constrained. On the other hand, this work can not put a constraint on the power spectrum on smaller scales, $k_{\m{s}}\gtrsim 60{\m{Mpc}^{-1}}$. 
This is because such diminutive UCMHs have no power to change the global ionization evolution.

\section{Conclusion}
In this work, we investigated the effect of Pop.~III stars in UCMHs on the cosmic ionization history using Planck observation data. 
Depending on the scale or the amplitude of the additional spike-type power spectrum, UCMHs could be formed in an earlier epoch compared to the standard halo formation scenario. 
Although the high-redshift astrophysics is not understood yet, UCMHs might host the Pop.~III stars like the standard halos. 
Such Pop.~III stars would emit the ionizing photon during their main sequence, and as a result, facilitate the cosmic reionization in high redshifts.

Since the high-redshift reionization can affect the CMB anisotropies, the CMB anisotropy measurement allows us to test the effect of UCMH Pop.~III stars. 
In order to investigate the effect, we have implemented the MCMC analysis with the latest Planck observation data for the reionization model including the effect of UCMH Pop.~III stars.
In this work, we employ the conventional ``tanh"-type reionization model which represents the contribution from the first galaxies and Pop.~II stars as the main sources of ionization photons.
Then, we investigated the following three parameters which compose our reionization model: (1)the mass variance on the scale $k_{\m{s}}$ represented by $\mathcal{A}_{\m{mat,0}}$ which relates to the amplitude of the additional power spectrum, (2)the standard reionization parameter $z_{\m{reio}}$ which controls the conventional ``tanh"-type reionization, and (3) initial mass models of UCMHs in the range of $10^{7}M_{\odot}<M_{\m{i}}<10^{9}M_{\odot}$ corresponding to the spike scale $k_{\m{s}}$ of the additional power spectrum.
We performed separate MCMCs for several UCMH initial mass models to explore the other parameters of $\mathcal{A}_{\m{mat,0}}$ and $z_{\m{reio}}$.

We have found that when $M_{\m{i}} < 10^{8}\m{M}_{\odot}$, the UCMH Pop.~III stars contribution is totally subdominant no matter how large $\mathcal{A}_{\m{mat,0}}$ is, and the constraint on our reionization model is almost same as the constraint without the UCMH Pop.~III stars.
However, as the initial mass becomes larger, $M_{\m{i}} > 10^{8}\m{M}_{\odot}$, UCMH Pop.~III stars gradually affect the Thomson scattering optical depth of the CMB and the reionization history. Then the ``tanh"-type reionization would be delayed to compensate for the early reionization due to the UCMH Pop.~III stars. Once the UCMHs mass becomes larger than the minimum mass to host Pop.~III stars, their contribution drastically increases. 
In that case, the decrement of $z_{\m{reio}}$ can not cancel out the increase of $\tau$ coming from large $\mathcal{A}_{\m{mat,0}}$, and the constraint on the value of $\mathcal{A}_{\m{mat,0}}$ becomes stringent. 
From the constraint on the value of $\mathcal{A}_{\m{mat,0}}$, one can put a constraint on the amplitude of the additional spike-type power spectrum, $\mathcal{A}_{\zeta}^{\m{add}}$. 
As one assumes that the standard ``tanh"-type reionization occurs by $z=3$, one can obtain the constraint, $\mathcal{A}_{\zeta}^{\m{add}}\lesssim 10^{-8}$ in the scales, $k\lesssim 50\m{Mpc}^{-1}$.

Before finishing the conclusion, we mention some caveats for this work.
In this paper, we fixed the unknown astrophysical properties of Pop.~III stars and UCMHs following the previous work~\cite{2020PhRvD.101l3518I}.
We need to implement simulations following the formation of UCMHs and the Pop.~III stars to check the validity of these properties.
We leave this possibility for future work.

We also need to mention the formation of Pop.~II stars and the connection between the Pop.~III stars formation and the Pop.~II stars formation.
In order to take this into account properly, we need to consider the lifetime of Pop.~III stars, their fate to produce heavier elements, and the increment of cosmic metalicity which leads to Pop.~II stars' formation. We also leave this possibility for future work.



\acknowledgments
We would like to thank Hiroyuki Tashiro for the helpful discussions.
K.T.A is supported by the Japan Society for the Promotion of Science~(JSPS) KAKENHI No.~JP20J22260.

\bibliography{article}
\end{document}